# Investigation of beam self-polarization in the future $e^+e^-$ circular collider

E. Gianfelice-Wendt

*Fermilab, Batavia, Illinois 60510, USA*

The use of resonant depolarization has been suggested for precise beam energy measurements (better than 100 keV) in the $e^+e^-$ Future Circular Collider (FCC-$e^+e^-$) for $Z$ and $WW$ physics at 45 and 80 GeV beam energy respectively. Longitudinal beam polarization would benefit the $Z$ peak physics program; however it is not essential and therefore it will be not investigated here. In this paper the possibility of self-polarized leptons is considered. Preliminary results of simulations in presence of quadrupole misalignments and beam position monitors (BPMs) errors for a simplified FCC-$e^+e^-$ ring are presented.

## I. INTRODUCTION

A hadron collider of 100 km circumference and 100 TeV in the center of mass is under study by the international physics community as next energy frontier Future Circular Collider (FCC). The same tunnel could host first a lepton collider, FCC-$e^+e^-$, with beam energy ranging between 45 and 175 GeV, cost-effectively combining the Triple-Large Electron-Positron collider and Very High Energy Large Hadron Collider proposals [1].

Accurate energy determination is a fundamental ingredient of precise electroweak measurements. Resonant depolarization has been used with this purpose for the first time at VEPP-2M [2] and more recently at Large Electron Positron collider (LEP) [3].

The aim of this paper is to present results of preliminary investigations on the level of polarization attainable in FCC-$e^+e^-$. The energy calibration scenario for FCC-$e^+e^-$ is described in [4].

The principle behind resonant depolarization is that a vertically polarized beam excited through an oscillating horizontal magnetic field gets depolarized when the excitation frequency is in a given relationship with the beam energy.

A lepton beam may get vertically polarized in the guiding field of a storage ring through the Sokolov-Ternov effect [5]. In an ideally planar ring[1] the asymptotic polarization is $P_{ST} = 92.4\%$ and the polarization build-up rate is given by

$$\tau_p^{-1} = \frac{5\sqrt{3}}{8} \frac{r_e \gamma^5 \hbar}{m_0 C} \oint \frac{ds}{|\rho|^3}.$$

---

[1]With *ideally planar* it is meant that the design orbit is contained in a plane and the magnets are perfectly aligned. To be more precise we shall also exclude the presence of dipole fields pointing in the direction opposite to the guiding field.

In actual ring accelerators there are not only dipoles. Quadrupoles for instance are needed for beam focusing. When a particle emits a photon it starts to perform synchro-betatron oscillations around the machine actual closed orbit experiencing extra, possibly non vertical, fields. The expectation value $\vec{S}$ of the spin operator obeys to the Thomas-Bargmann-Michel-Telegdi (Thomas-BMT) equation [6,7]

$$\frac{d\vec{S}}{dt} = \vec{\Omega} \times \vec{S} \qquad (1)$$

$\vec{\Omega}$ depends on machine azimuth and phase space position, $\vec{u}$. In the laboratory frame and MKS units it is given by

$$\vec{\Omega}(\vec{u};s) = -\frac{e}{m_0}\left[\left(a+\frac{1}{\gamma}\right)\vec{B} - \frac{a\gamma}{\gamma+1}\vec{\beta}\cdot\vec{B}\vec{\beta} - \left(a+\frac{1}{\gamma+1}\right)\vec{\beta}\times\frac{\vec{E}}{c}\right]$$

with $\vec{\beta} \equiv \vec{v}/c$ and $a = (g-2)/2 = 0.0011597$ (for $e^{\pm}$).

In a planar machine the *periodic* solution, $\hat{n}_0$, to Eq. (1) is vertical and, neglecting the electric field, the number of spin precessions around $\hat{n}_0$ per turn, the naive "spin tune," in the rotating frame is $a\gamma$. Photon emission results in a randomization of the particle spin directions (*spin diffusion*). Polarization will be therefore the result of the competing process, the Sokolov-Ternov effect and the spin diffusion caused by stochastic photon emission.

Using a semiclassical approach, Derbenev and Kondratenko [8] found that the polarization is oriented along $\hat{n}_0$ and its asymptotic value is

$$P_{DK} = P_{ST} \frac{\oint ds \langle \frac{1}{|\rho|^3} \hat{b}\cdot(\hat{n}-\frac{\partial\hat{n}}{\partial\delta})\rangle}{\oint ds \langle \frac{1}{|\rho|^3}[1-\frac{2}{9}(\hat{n}\cdot\hat{v})^2+\frac{11}{18}(\frac{\partial\hat{n}}{\partial\delta})^2]\rangle} \qquad (2)$$

with $\hat{b} \equiv \hat{v} \times \dot{\hat{v}}/|\dot{\hat{v}}|$ and $\delta \equiv \delta E/E$. There have been some disputes about the meaning of the quantity $\hat{n}$ and its

derivative in the original paper. Nowadays $\hat{n}$ is understood as an *invariant spin field* [9], i.e. a solution of Eq. (1) satisfying the condition $\hat{n}(\vec{u}; s) = \hat{n}(\vec{u}; s + C)$, $C$ being the machine length. The $\langle \rangle$ brackets indicate averages over the phase space. The term $\partial \hat{n}/\partial \delta$ quantifies the depolarizing effects resulting from the trajectory perturbations due to photon emission.

The corresponding polarization rate is

$$\tau_p^{-1} = P_{ST} \frac{r_e \gamma^5 \hbar}{m_0 C} \oint \left\langle \frac{1}{|\rho|^3} \left[ 1 - \frac{2}{9}(\hat{n} \cdot \hat{v})^2 + \frac{11}{18}\left(\frac{\partial \hat{n}}{\partial \delta}\right)^2 \right] \right\rangle. \quad (3)$$

In a perfectly planar machine $\partial \hat{n}/\partial \delta = 0$ and $P_{DK} = P_{ST}$. In the presence of quadrupole vertical misalignments (and/or spin rotators) $\partial \hat{n}/\partial \delta \neq 0$ and it is particularly large when spin and orbital motions are in resonance

$$\nu_{\text{spin}} \pm m Q_x \pm n Q_y \pm p Q_s = \text{integer}. \quad (4)$$

For FCC-$e^+e^-$ with $\rho \simeq 10424$ m, fixed by the maximum attainable dipole field for the hadron collider, the polarization time at 45 and 80 GeV are 256 and 14 hours, respectively.

Here it is assumed that beam polarization of about 10% is sufficient for an accurate depolarization measurement. The time, $\tau_{10\%}$, needed for the beam to reach this polarization level is given by

$$\tau_{10\%} = -\tau_p \times \ln(1 - 0.1/P_\infty).$$

At 80 GeV it is $\tau_{10\%} = 1.6$ hours, but $\tau_{10\%} = 29$ hours at 45 GeV.

At low energy the polarization time may be reduced by introducing properly designed wiggler magnets i.e. a sequence of vertical dipole fields, $\vec{B}_w$, with alternating signs.

FCC-$e^+e^-$ maximum synchrotron radiation power is set to 50 MW per beam and the beam current at the various energies as been scaled accordingly. This limits the integrated wiggler strength. Moreover the wiggler increases the beam energy spread for which the effect on polarization must be investigated.

At 80 GeV wigglers are not needed. However the energy dependence of the spin motion makes the attainable polarization level more sensitive to machine errors.

In this paper studies are presented for a FCC-$e^+e^-$ "toy" ring consisting of 1064 Focusing-Open space–Defocusing-Open space (FODO) cells with dipole bending radius of $\rho_d \simeq 10424$ m and four dispersion free regions for accommodating wigglers. The fractional betatron tunes are $Q_x = 0.128$, $Q_y = 0.208$. $60°/60°$ and $90°/90°$ FODO cells have been considered.

The choice of using a simplified lattice is justified by the fact that the interaction regions (IRs) are not yet finalized and, more importantly, by the wish to disentangle the general problem from those arising from the peculiarities of the IR layout.

Evaluation of Eqs. (2) and (3) requires a precise knowledge of $\hat{n}(\vec{u}; s)$. For high energy large rings in presence of machine errors semianalytical approaches have either convergence problems or require very large computing power.

Here MAD-X is used for simulating quadrupole misalignments and closed orbit correction. It is assumed that there is a beam position monitor (BPM) and a vertical corrector dipole close to each vertical focusing quadrupole. The lattice with errors and corrections is dumped to a file which can be read by the SITROS package [10] used for polarization calculations. The package includes a module where orbit and spin motion are linearized, and a Monte-Carlo tracking with 2nd order orbit motion and nonlinear spin motion.

The original SITROS version has been improved to make it suitable for HERA-e polarization studies [11,12]. In the version used here the number of magnets where radiation is emitted has been increased to 2200.

When needed, the deviation, $\delta \hat{n}_0(s)$, of the periodic solution to Eq. (1) on the actual closed orbit from the nominal one, has been corrected by using vertical closed orbit bumps ("harmonic bumps") [13] in the arc cells. The effect of such bumps on $\hat{n}_0(s)$ depends on energy and cell optics and therefore the bumps configuration needs in general to be reoptimized if the beam energy and/or optics are changed.

## II. ENERGY SPREAD AND SYNCHROTRON TUNE

Lower order resonances are expected to dominate the spin dynamics. From Eq. (4) one can argue that small values of the orbital tunes are convenient for getting a low order resonance free "plateau" in between two zeroth order (or *imperfection*) resonances. This choice worked well for HERA-e at 27.5 GeV and $\sigma_E = 27$ MeV.

However the particle spin tune depends on the particle energy which oscillates with the synchrotron frequency. A large energy spread introduces a large modulation of spin and betatron tunes. In addition, the distance between two zeroth order resonances is 440 MeV independently from energy. It is to be expected that a large energy spread is harmful for polarization.

Indeed in [14,15] it is predicted that under the assumption resonances are well separated, the energy spread enhances the strength of the synchrotron side bands of the lower order resonances by the *"enhancement factor"*

$$\xi = \left(\frac{a\gamma}{Q_s} \frac{\sigma_E}{E}\right)^2. \quad (5)$$

At very high energy and in presence of unavoidable alignment and/or field errors the Sokolov-Ternov effect is overwhelmed by the spin diffusion. However in [16] a polarization resurrection is predicted if the condition

$$\frac{(a\gamma)^2 T_{\text{rev}}}{\tau_p Q_s^3} \ll 1 \quad (6)$$

is satisfied. Equations (5) and (6) suggest that large values of $Q_s$ may be desirable at high energy and/or large $\sigma_E$. Obtaining larger synchrotron tunes while keeping the rf voltage to reasonable values could require increasing the momentum compaction factor thus adding to the complexity of the optics.

At LEP no polarization was observed at beam energy above 65 GeV [17]. This was understood as a consequence of the increased energy spread. Extrapolating from LEP, an upper limit $\sigma_E = 50$ MeV was set for FCC. As we will see, that extrapolation is not fully correct; however considerations of radiation losses limit anyway the strength of the wiggler field.

In the following $Q_s$ has been fixed to $\simeq 0.1$. With $f_{RF} = 400$ MHz the needed voltage for the 60 degrees FODO is 0.9 GV at 45 GeV and 1.6 GV at 80 GeV, while for the 90 degrees FODO it is 1.8 GV at 45 GeV and 3 GV at 80 GeV. The voltage has been recomputed for keeping $Q_s$ constant when wigglers are powered on.

## III. POLARIZATION IN PRESENCE OF WIGGLER MAGNETS

At 45 GeV the polarization time may be reduced by using wigglers. In order not to perturb the design orbit the conditions

$$\int_{\text{wig}} ds\, B_w = 0 \quad \text{and} \quad \int_{\text{wig}} ds\, s B_w = 0$$

must be fulfilled. A symmetric wiggler with vanishing field integral fulfills both conditions.

In the presence of wigglers the polarization rate and the asymptotic polarization may be written as

$$\tau_p^{-1} = F\gamma^5 \left[ \int_{\text{dip}} \frac{ds}{|\rho_d|^3} + \int_{\text{wig}} \frac{ds}{|\rho_w|^3} \right] \quad F \equiv \frac{5\sqrt{3}}{8} \frac{r_e \hbar}{m_0 C}$$

$$P_\infty = \frac{8}{5\sqrt{3}} \frac{\oint ds\, \frac{\hat{B}\cdot\hat{n}_0}{|\rho|^3}}{\oint ds\, \frac{1}{|\rho|^3}} \propto \tau_p \left[ \int_{\text{dip}} ds \frac{\hat{B}_d \cdot \hat{n}_0}{|\rho_d|^3} + \int_{\text{wig}} ds \frac{\hat{B}_w \cdot \hat{n}_0}{|\rho_w|^3} \right]$$

which may be obtained from Derbenev-Kondratenko expressions for a perfectly planar machine where $\partial \hat{n}/\partial\delta = 0$ and $\hat{n}\cdot\hat{v}=0$. In order to reduce the polarization time significantly $|B_w|$ must be large and $\int_{\text{wig}} ds B_w^3$ must be large too in order to keep $P_\infty$ large.

The use of wiggler magnets for reducing the polarization time was first proposed for LEP [18]. Their schematic layout is shown in Fig. 1.

A similar layout has been considered here.

As for polarization the main concern is the dynamics in the vertical plane, we will stick here to the LEP-like design keeping in mind that a larger number of poles may be more convenient not to increase the horizontal emittance [19,20].

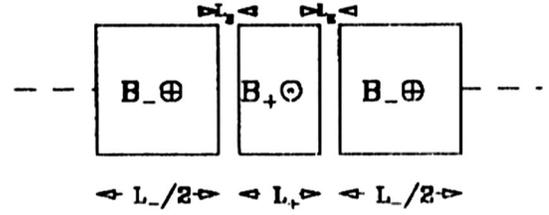

FIG. 1. LEP polarization wiggler (figure from [18]).

Since $\hat{n}_0 \equiv \hat{y}$ in a perfectly planar ring and the field is piecewise constant in the wiggler, the asymptotic polarization and the polarization build-up rate write

$$P_\infty = \frac{8F\gamma^5}{5\sqrt{3}} \tau_p \left[ \int_{\text{dip}} ds \frac{\hat{B}_d \cdot \hat{n}_0}{|\rho_d|^3} + \frac{L^+}{|\rho^+|^3}\left(1 - \frac{1}{N^2}\right)\right]$$

$$\tau_p^{-1} = F\gamma^5 \left[ \int_{\text{dip}} \frac{ds}{|\rho_d|^3} + \int_{\text{wig}} \frac{ds}{|\rho_w|^3} \right]$$

$$= F\gamma^5 \left[ \int_{\text{dip}} \frac{ds}{|\rho_d|^3} + \frac{L^+}{|\rho^+|^3}\left(1 + \frac{1}{N^2}\right)\right]$$

where $N \equiv L^-/L^+ = B^+/B^-$.

The particle energy lost per turn and the energy spread are

$$U_{\text{loss}} = \frac{C_\gamma E^4}{2\pi} \oint \frac{ds}{\rho^2} \quad (\sigma_E/E)^2 = \frac{C_q}{J_\epsilon}\gamma^2 \oint \frac{ds}{|\rho|^3} \bigg/ \oint \frac{ds}{\rho^2}.$$

The presence of wigglers increases $U_{\text{loss}}$ and $\sigma_E/E$.

Is it possible to optimize the wiggler layout/number for minimizing the increment of $U_{\text{loss}}$ and $\sigma_E/E$?

The generally valid relationship

$$(\sigma_E/E)^2 = \frac{C_q C_\gamma E^4}{2\pi J_\epsilon F\gamma^3} \frac{1}{\tau_p U_{\text{loss}}}$$

shows that a small $\tau_p$ is at price of a higher $U_{\text{loss}}$ and/or $\sigma_E$. Figure 2 shows the relevant parameter values vs. polarization time in presence of one wiggler for $E = 45$ GeV and different values of $N$, $L_+$ being fixed to 1.3 m. Figure 3 shows the same parameters vs. polarization time for different numbers of wiggler units at $E = 45$ GeV with $N = 6$ and $L^+ = 1.3$ m. There is no "miraculous" set of wiggler parameters, all cases with $N > 2$ being almost equivalent.

In the following we will consider at 45 GeV the case of four wigglers with $N = 6$, $L^+ = 1.3$ m and $B^+ = 0.7$ T, corresponding to $\tau_{10\%} = 2.9$ h and to an increase of the energy spread from 18 MeV to 50 MeV. Dispersion and $\beta$ functions in the wiggler section with $B^+ = 0.7$ T are shown in Fig. 4. The horizontal emittance increases from $0.8 \times 10^{-3}$ μm to $1.0 \times 10^{-3}$ μm when the wiggler field is on.

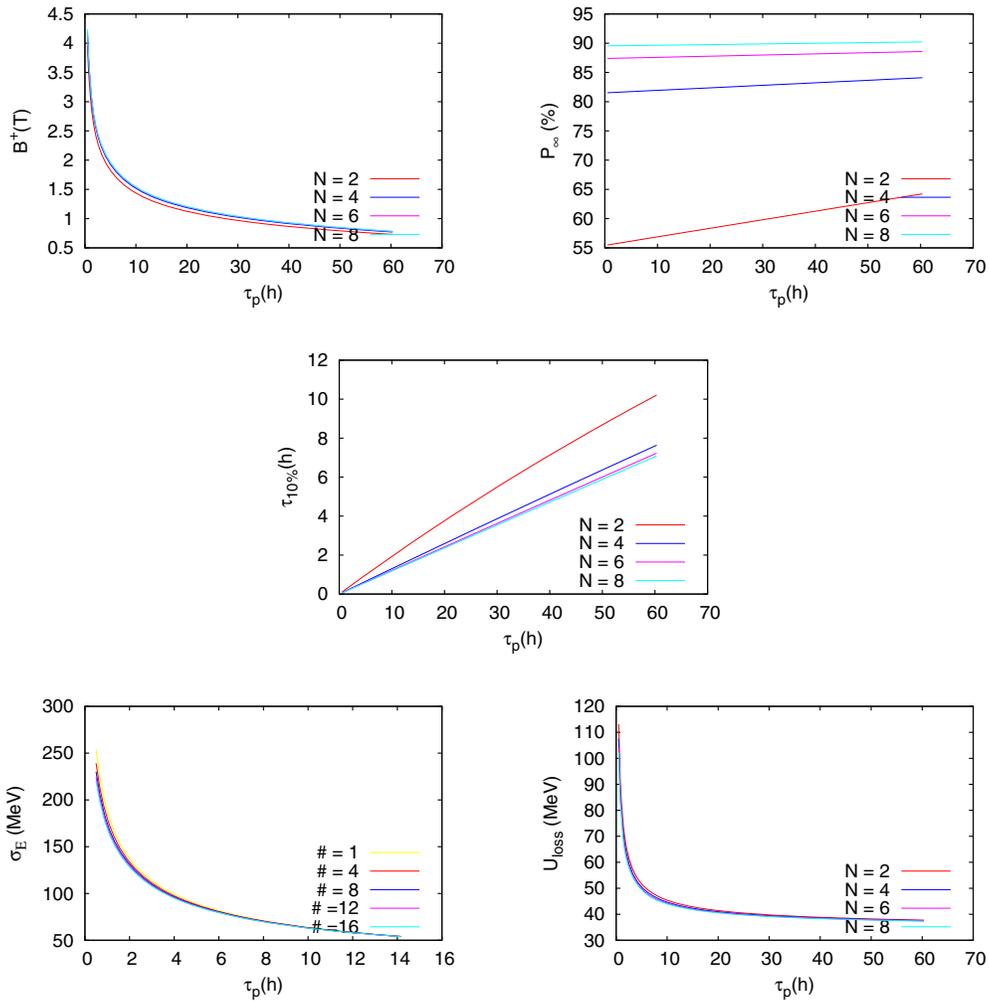

FIG. 2. Parameter values vs. polarization time for one wiggler at 45 GeV for various values of $N$. Top left: $B^+$. Top right: Asymptotic polarization. Center: $\tau_{10\%}$. Bottom left: Energy spread. Bottom right: Energy lost per turn.

## IV. SIMULATIONS FOR THE 60°/60° FODO

### A. 60°/60° FODO—45 GeV beam energy

According to Eq. (2) spin diffusion appears when the machine is not perfectly planar. The presence of wigglers alone does not introduce any spin diffusion.

The main source of spin diffusion is usually the vertical misalignment of quadrupoles which makes $\hat{n}(\vec{u};s)$ not homogeneous. Simulations in the presence of realistic vertical quadrupole misalignments are needed to assess their effect on polarization and the role of the increased energy spread due to the wiggler, as well as to evaluate the efficacy of corrective strategies.

Polarization vs. $a\gamma$ for 200 $\mu$m rms vertical quadrupole offset, $\delta_y^Q$, is shown in Fig. 5 after the orbit has been corrected by using *all* 1096 correctors through singular value decomposition (SVD) for one particular seed. The rms closed orbit is decreased from 8 mm to 0.05 mm and $|\delta\hat{n}_0|_{rms}$ is decreased from 26.4 mrad to 0.3 mrad. The correction is almost *local* and polarization, which was close to zero before correction, is restored despite wiggler presence. The nonlinear computation (the blue crosses in Fig. 5) is very close to the linear one (cyan line). However also errors on the BPMs system must be considered. Figure 6 left shows polarization when 200 $\mu$m BPMs rms vertical displacement, $\delta_y^M$, and 10% random calibration error is added. In this case after correction the rms closed orbit is 0.8 mm and $|\delta\hat{n}_0|_{rms} = 3.9$ mrad. It is possible by tuning 8 "harmonic bumps" in the arc cells to further decrease $|\delta\hat{n}_0|_{rms}$ by a factor two. The resulting polarization is shown in Fig. 6 right. Figure 7 shows polarization vs. $a\gamma$ for the same machine after increasing $B^+$ from 0.7 T to 3.9 T, which of course leaves orbit and $\delta\hat{n}_0$ unchanged. The corresponding $\sigma_E$ is 247 MeV, $U_{\text{loss}} = 278$ MeV/turn and $\tau_{10\%} = 1$ minute. The effect of the energy spread becomes evident when the spin motion is not linearized confirming the role of the energy spread. However due to the relatively small $\delta\hat{n}_0$ achieved the level of polarization would be sufficient for energy calibration if one could afford the large energy loss.

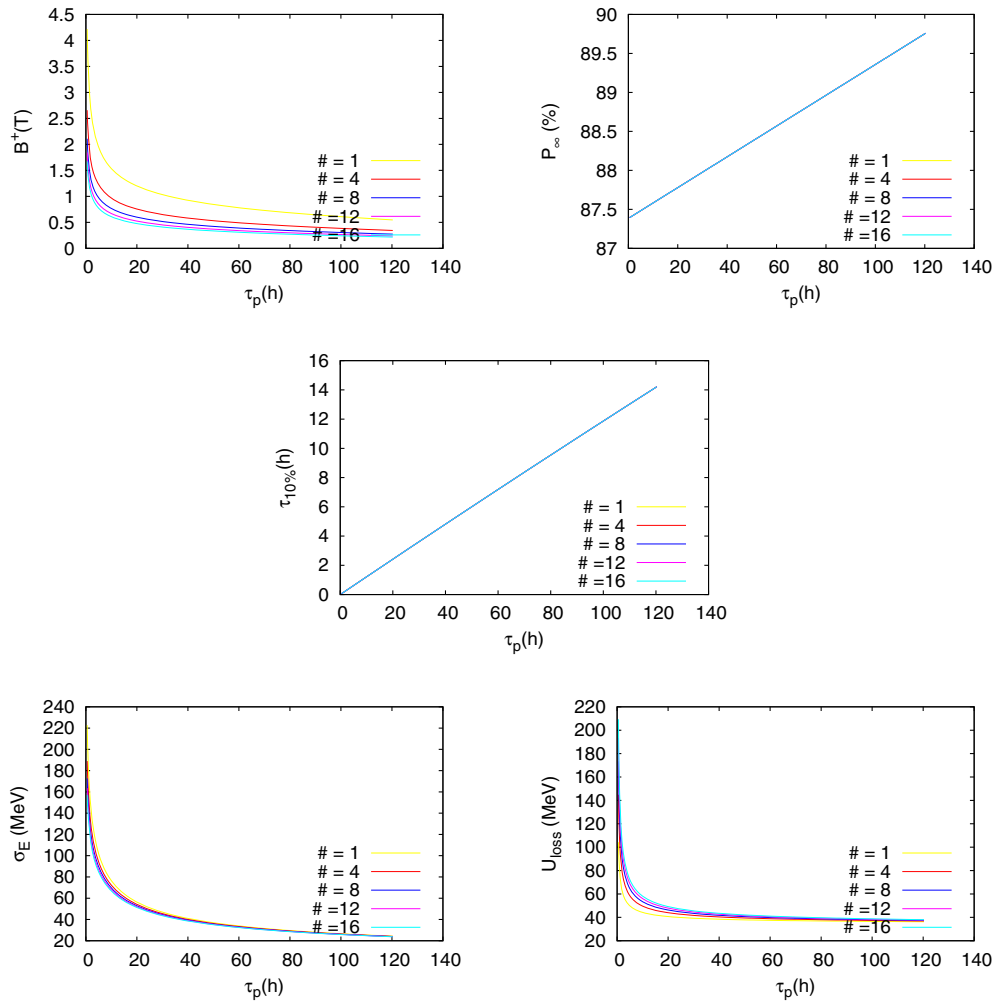

FIG. 3. Parameter values vs. polarization time for different numbers of wigglers units at 45 GeV with $N = 6$ and $L^+ = 1.3$ m. Top left: $B^+$. Top right: Asymptotic polarization. Center: $\tau_{10\%}$. Bottom left: Energy spread. Bottom right: Energy lost per turn.

The polarization achieved for the same machine with a less aggressive closed orbit correction using a limited number of correctors followed by $\delta \hat{n}_0$ correction is shown in Fig. 8. The 110 vertical correctors for the closed orbit correction are selected through the well-known MICADO algorithm.

### B. 60°/60° FODO—80 GeV beam energy

At 80 GeV wigglers are not needed. However the larger synchrotron radiation emitted by the particles increases spin diffusion and energy spread (54 MeV, comparable to the 45 GeV case with wigglers). In addition the same quadrupole displacement distribution will produce a larger $\delta \hat{n}_0$.

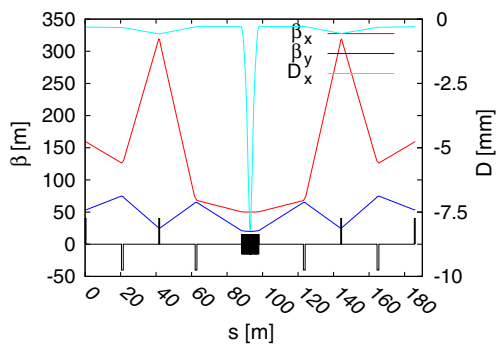

FIG. 4. Twiss and dispersion functions in the wiggler section with $B^+ = 0.7$ T. The Twiss functions are practically unchanged; the peak $|D_x|$ (and orbit deviation) is 8 mm.

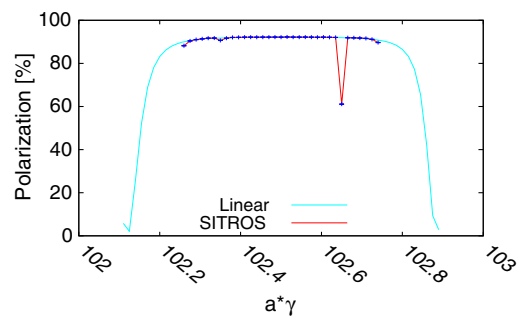

FIG. 5. Polarization vs. spin tune after orbit is corrected by using all 1096 correctors with $\delta_y^Q = 200$ μm, w/o BPMs errors and in presence of 4 wigglers with $B^+ = 0.7$ T. 60°/60° FODO.

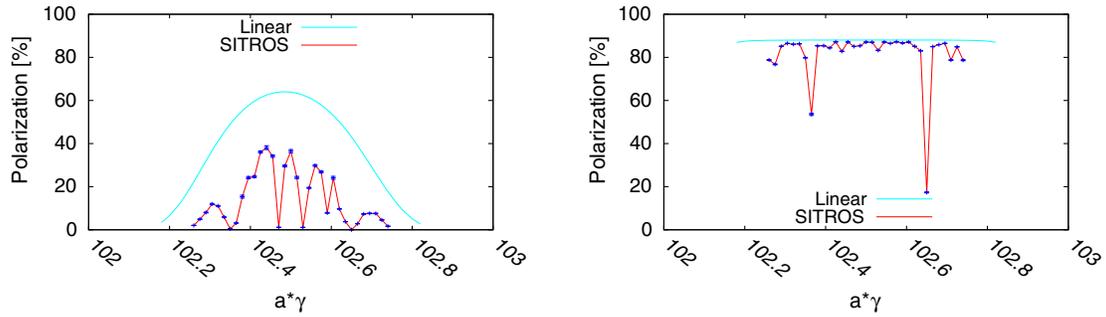

FIG. 6. 60°/60° FODO. Left: Polarization vs. spin tune after orbit is corrected with 1096 correctors with $\delta_y^Q = \delta_y^M = 200$ $\mu$m and 10% BPM calibration error; 4 wigglers with $B^+ = 0.7$ T. Right: adding the correction of $\delta \hat{n}_0$ through harmonic bumps.

Figure 9 left shows polarization for the same seed used at 45 GeV and w/o BPMs errors. The SVD reduces $|\delta \hat{n}_0|_{rms}$ to 3 mrad (it was 0.3 mrad at 45 GeV). Comparing Fig. 9 left with the equivalent case at 45 GeV (Fig. 5) there is a clear deterioration. The correction of $\delta \hat{n}_0$ which was not necessary at 45 GeV in absence of BPMs errors is now needed to improve the polarization level (Fig. 9 right). Adding BPMs random misalignments ($\delta_y^M = 200$ $\mu$m) and 10% calibration error it became evident that the $\delta \hat{n}_0(s)$ correction, which was adequate at 45 GeV, must be improved at 80 GeV. In fact the orbit correction, which brings the rms orbit down to 0.8 mm, leaves $|\delta \hat{n}_0|_{rms} = 19.9$ mrad with almost no polarization. The large bump amplitudes needed to correct $\delta \hat{n}_0$ (see Fig. 10) have a large impact on the emittance ratio, $\epsilon_y/\epsilon_x$, which increases from 0.2% to 3%. As a consequence even the linearized calculation predicts little polarization (see Fig. 11 left).

Instead of using orbit bumps involving 3 correctors, one can use 5 correctors and ask that the vertical dispersion is unchanged outside the bump region. By tuning 8 such bumps it is possible to achieve $|\delta \hat{n}_0|_{rms} = 9.7$ mrad without affecting the vertical emittance. The resulting polarization is shown in Fig. 11 right.

At 80 GeV, $\delta \hat{n}_0$ due to the same misalignments is larger than at 45 GeV and although the energy spread is the *same* as at 45 GeV with wigglers, polarization is lower.

A more careful search between 3-correctors harmonic bumps allows identification of a configuration which could correct $\delta \hat{n}_0$ with very small beam offsets and no emittance growth. The resulting polarization was similar to the 5-correctors harmonic bumps case.

Although "academic" for FCC, it is interesting to investigate the importance of the synchrotron tune. By increasing $Q_S$ from 0.1 to 0.3 the enhancement factor $\xi$ [Eq. (5)] is reduced from 1.5 to 0.17 while the ratio on the l.h.s. of Eq. (6) is reduced from 0.2 to $8 \times 10^{-3}$. Polarization for the same machine and $Q_s = 0.3$ (11 GV RF voltage) is shown in Fig. 12 left. Synchrotron sidebands move to the center with no evident gain. Increasing the synchrotron tune to 0.9, the location of the sidebands is as for $Q_s = 0.1$; now a large improvement is observed (see Fig. 12 right). For FCC at

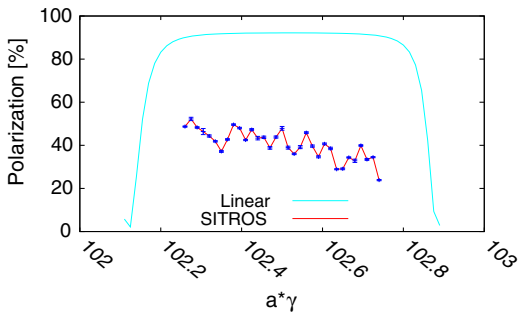

FIG. 7. Polarization vs. spin tune after orbit is corrected with 1096 correctors in presence of BPMs errors with $B^+ = 3.9$ T.

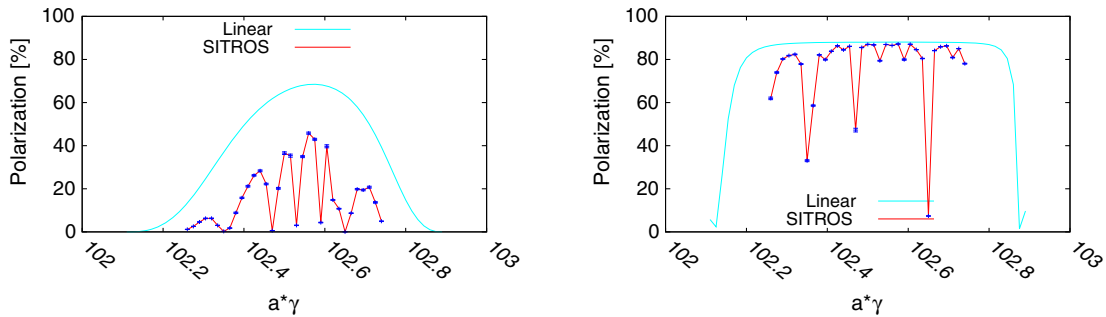

FIG. 8. 60°/60° FODO. Left: Polarization vs. spin tune after orbit is corrected with 110 correctors with $\delta_y^Q = \delta_y^M = 200$ $\mu$m and 10% BPM calibration error; 4 wigglers with $B^+ = 0.7$ T. Right: adding the correction of $\delta \hat{n}_0$ through orbit bumps.

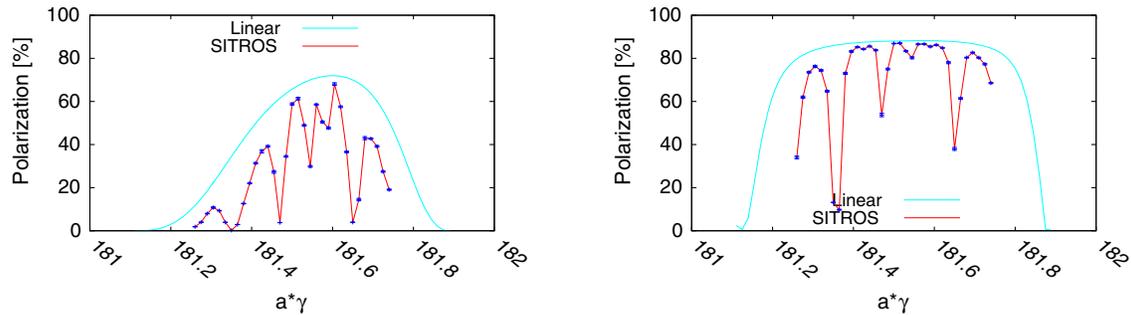

FIG. 9. 60°/60° FODO. Left: Polarization vs. spin tune after orbit is corrected with 1096 correctors with $\delta_y^Q = 200$ μm and no BPMs errors. Right: adding the correction of $\delta\hat{n}_0$ through orbit bumps.

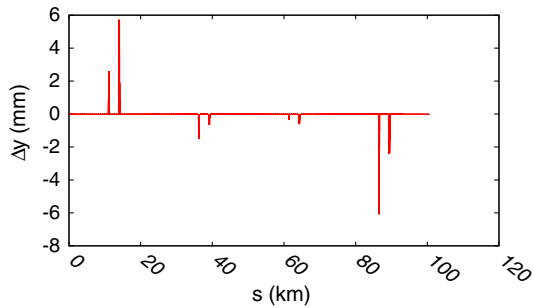

FIG. 10. Vertical closed orbit change due to the 8 3-correctors harmonic bumps for the 80 GeV case.

80 GeV such large synchrotron tune would require a sum voltage of 91 GV.

## V. SIMULATIONS FOR THE 90°/90° FODO

The current design of the FCC-$e^+e^-$ arcs is based on 90°/90° FODO cells. The nominal horizontal emittance is therefore smaller than in the previous cases. This may be an advantage for polarization once also other sources of errors are included, for instance those introducing betatron coupling. However the misalignment of stronger quadrupoles will cause a larger perturbation of the orbit. Some simulations for the 90°/90° FODO are presented here.

### A. 90°/90° FODO—45 GeV beam energy

Introducing the same distribution of errors the rms vertical closed orbit is 14 mm, instead of 8 mm, reduced to 0.08 mm by correcting with all 1096 correctors through a SVD. The corresponding $|\delta\hat{n}_0|_{rms}$ is 32.5 mrad and 0.3 mrad before and after orbit correction, respectively. In the presence of BPMs errors the rms vertical closed orbit after correction is 1.4 mm and $|\delta\hat{n}_0|_{rms} = 10$ mrad. The relatively large residual orbit is due mainly to the BPMs calibration error and therefore targeting the vertical dispersion, measured through orbit *variations* in response to rf frequency changes, instead of the vertical closed orbit would not help in improving the correction. Polarization after orbit correction w/o and with BPMs errors is shown in Fig. 13. In the latter case by using 8 vertical closed orbit bumps it is possible to correct $|\delta\hat{n}_0|_{rms}$ down to 6.2 mrad. The corresponding polarization is shown in Fig. 14 left.

### B. 90°/90° FODO—80 GeV beam energy

After orbit correction with BPMs errors is $|\delta\hat{n}_0|_{rms} = 35$ mrad reduced to 14 mrad by tuning 8 3-correctors harmonic bumps. The corresponding polarization is shown in Fig. 14 right.

Comparing Fig. 14 right with Fig. 11 right (60°/60° FODO) we notice that the level of polarization as computed by SITROS are not much different, however the linear calculation foresees a larger polarization for the 90°/90°

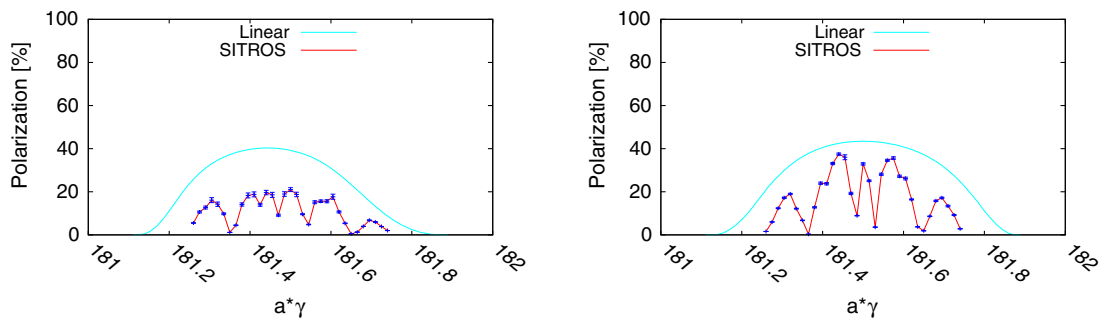

FIG. 11. 60°/60° FODO. Polarization vs. spin tune after orbit is corrected with SVD in presence of BPMs errors. $\delta\hat{n}_0$ is corrected by 8 3-correctors harmonic bumps (left) or by 8 dispersion-free 5-correctors harmonic bumps (right).

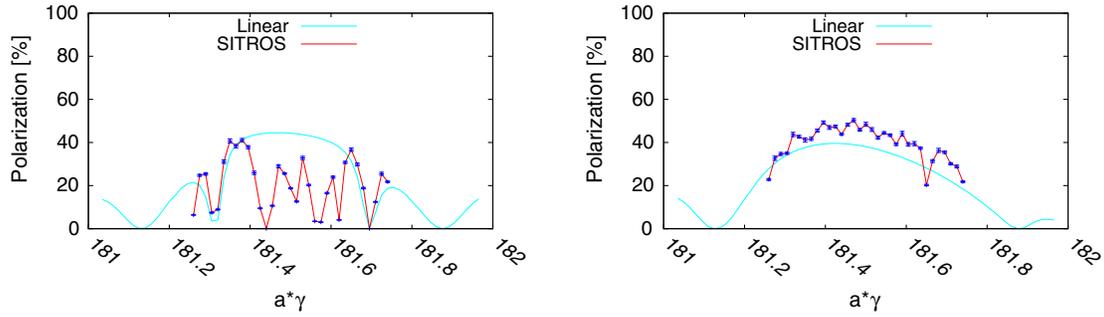

FIG. 12. Polarization vs. spin tune after orbit is corrected with SVD in presence of BPMs errors and $\delta\hat{n}_0$ is corrected with 5-correctors harmonic bumps. Left: $Q_s = 0.3$. Right: $Q_s = 0.9$.

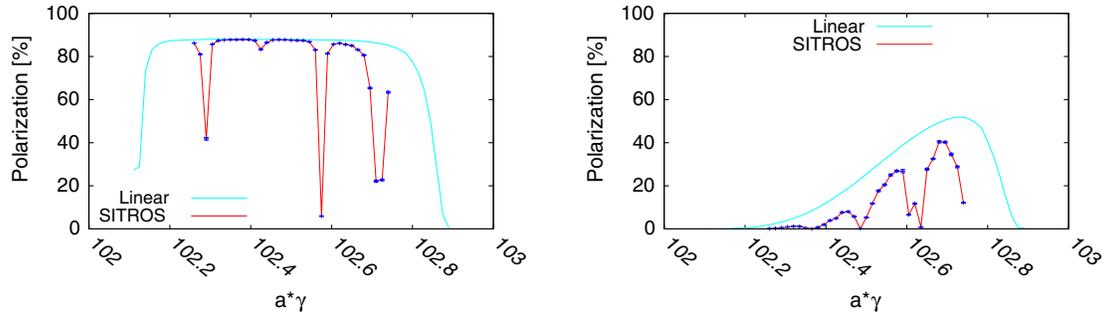

FIG. 13. 90°/90° FODO. Left: Polarization vs. spin tune after orbit is corrected with 1096 correctors with $\delta_y^Q = 200$ $\mu$m and w/o BPMs errors; 4 wigglers with $B^+ = 0.7$ T. Right: adding the BPMs errors.

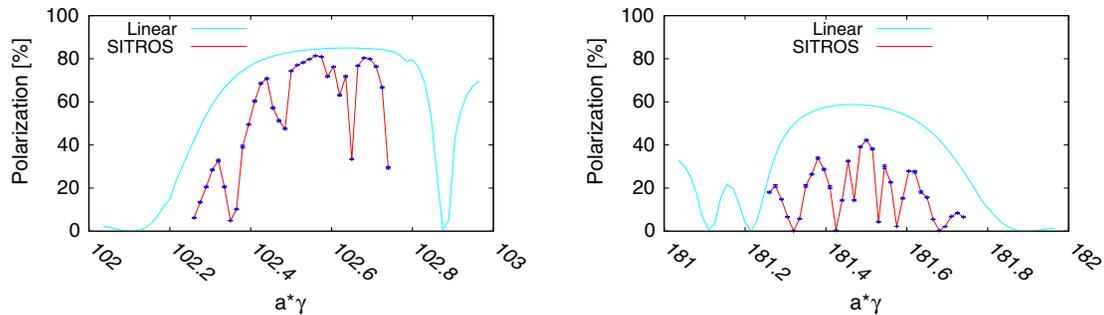

FIG. 14. 90°/90° FODO. Left: Polarization vs. spin tune after orbit and $\delta\hat{n}_0$ correction with $\delta_y^Q = \delta_y^M = 200$ $\mu$m and 10% BPMs calibration error; 4 wigglers with $B^+ = 0.7$ T (45 GeV). Right: Polarization vs. spin tune after orbit and $\delta\hat{n}_0$ correction with $\delta_y^Q = \delta_y^M = 200$ $\mu$m and 10% BPMs calibration error, no wigglers (80 GeV).

FODO. This may indicate that the correction for the 60°/60° FODO case at 80 GeV could be improved or simply that more statistics is needed.

## VI. CONCLUSIONS AND OUTLOOKS

Preliminary feasibility studies of polarization for energy calibration at FCC-$e^+e^-$ for 45 and 80 GeV beam energy have been presented.

At low energy the large bending radius requires wigglers for reducing the polarization time.

In the presence of errors, in particular the vertical misalignment of quadrupoles, depolarizing resonances appear. Synchrotron sidebands become more dangerous with increasing energy spread. Their importance can be quantified only by nonlinear calculations.

Maintaining acceptable level of polarization calls for well planned correction schemes, in particular at 80 GeV where $\delta\hat{n}_0$, energy spread and enhancement factor $\xi$ are larger. Here we have considered that there is a BPM and a corrector close to each vertical focusing quadrupole. This layout, adopted for instance in the arcs of HERA-e, is definitely recommended for FCC-$e^+e^-$ both for luminosity and polarization operation, in particular in the case 90° FODO cells are chosen. Moreover it opens the possibility of beam based

INVESTIGATION OF BEAM SELF-POLARIZATION … PHYS. REV. ACCEL. BEAMS 19, 101005 (2016)alignment techniques (see for instance [21] for its application to beam polarization).

The orbit has been corrected using either all or a limited number of correctors. The residual $\delta\hat{n}_0$ is corrected by using harmonic bumps.

Maintaining polarization for energy calibration seems feasible even in the presence of BPMs errors, the 80 GeV case being understandably more challenging.

The results here presented are instructive, however they refer to one single error seed and to a much simplified ring. Statistics is needed and the actual machine layout, including the interaction regions, will have to be considered in the near future.

SITROS results will be cross-checked by using SLICKTRACK [22].

Finally it must be noted that the relationship $\nu_{\text{spin}} = a\gamma$ on which energy calibration through resonant depolarization is based, holds for a purely planar ring. Electric and radial magnetic fields introduce deviations from this simple relationship and their effect must be assessed for evaluating whether the required precision (better than 100 keV) can be attained.

## ACKNOWLEDGMENTS

I would like to thank A. Blondel and M. Koratzinos for fruitful discussions. The support of the FCC Week organizing committee is warmly acknowledged. A special thanks goes to D. P. Barber who introduced me to the world of lepton polarization. Finally I thank B. C. Brown for carefully reading the manuscript. Work supported by Fermi Research Alliance LLC under DE-AC02-07CH11359 with the U.S. DOE.[1] M. Bicer et al. (TLEP Design Study Working Group Collaboration), First Look at the Physics Case of TLEP, J. High Energy Phys. 01 (2014) 164.

[2] Ya. S. Derbenev, A. M. Kondratenko, S. I. Serednyakov, A. N. Skrinsky, G. M. Tumaikin, and Yu. M. Shatunov, Accurate calibration of the beam energy in a storage ring based on measurement of spin precession frequency of polarized particles, Part. Accel. **10**, 177 (1980).

[3] R. W. Assmann et al. (LEP Energy Working Group Collaboration), Calibration of centre-of-mass energies at LEP1 for precise measurements of Z properties, Eur. Phys. J. C **6**, 187 (1998).

[4] M. Koratzinos, A. Blondel, E. Gianfelice-Wendt, and F. Zimmermann, FCC-ee: Energy calibration, arXiv:1506.00933.

[5] A. A. Sokolov and I. M. Ternov, On polarization and spin effects in the theory of synchrotron radiation, Sov. Phys. Dokl. **8**, 1203 (1964).

[6] L. H. Thomas, The kinematics of an electron with an axis, Philos. Mag. **3**, 1 (1927).

[7] V. Bargmann, L. Michel, and V. L. Telegdi, Precession of the Polarization of Particles Moving in a Homogeneous Electromagnetic Field, Phys. Rev. Lett. **2**, 435 (1959).

[8] Ya. S. Derbenev and A. M. Kondratenko, Zh. Eksp. Teor. Fiz. **64**, 1918 (1973) [Polarization kinematics of particles in storage rings, Sov. Phys. JETP **37**, 968 (1973)].

[9] G. H. Hoffstatter, M. Vogt, and D. P. Barber, Higher order effects in polarized proton dynamics, Phys. Rev. ST Accel. Beams **2**, 114001 (1999).

[10] J. Kewisch, Ph.D. thesis, Hamburg University, 1985.

[11] M. Boge, Ph.D. thesis, Hamburg University, 1994.

[12] G. Z. M. Berglund, Ph.D. thesis, Royal Institute of Technology, Stockholm, 2001.

[13] D. P. Barber et al., High spin polarization at the HERA electron storage ring, Nucl. Instrum. Methods Phys. Res., Sect. A **338**, 166 (1994).

[14] K. Yokoya, Effects of radiative diffusion on the spin flip in electron storage rings, Part. Accel. **14**, 39 (1983).

[15] S. R. Mane, Synchrotron sideband spin resonances in high-energy electron storage rings, Nucl. Instrum. Methods Phys. Res., Sect. A **292**, 52 (1990).

[16] Ya. S. Derbenev, A. M. Kondratenko, and A. N. Skrinsky, Radiative polarization at ultrahigh-energies, Part. Accel. **9**, 247 (1979).

[17] R. Assmann et al., Spin dynamics in LEP with 40-GeV–100-GeV beams, AIP Conf. Proc. **570**, 169 (2001).

[18] A. Blondel and J. M. Jowett, Wigglers for polarization, Conf. Proc. **C8711093**, 216 (1987).

[19] I. Koop (private communication).

[20] R. P. Walker, Advanced accelerator physics, in *Proceedings of the 5th Course of the CERN Accelerator School, Rhodos, Greece, 1993*, Vol. 1, 2 (CERN, 1994).

[21] D. P. Barber, R. Brinkmann, E. Gianfelice, T. Limberg, N. Meyners, P. Schuler, M. Spengos, and M. Boege, Application of a beam based alignment technique for optimizing the electron spin polarization at HERA, Conf. Proc. **C960610**, 439 (1996).

[22] D. P. Barber, Polarization in the eRHIC electron (positron) ring, in *Spin physics. Polarized electron sources and polarimeters. Proceedings, 16th International Symposium, SPIN 2004, Trieste, Italy, 2004, and Workshop, PESP 2004, Mainz, Germany, 2004* (World Scientific, Singapore, 2004), pp. 738–741.101005-9